# Investigation of Lasing in Highly Strained Germanium at the Crossover to Direct Band Gap.

F. T. Armand Pilon[1,5+], Y-M Niquet[2], J. Chretien[3], N. Pauc[3], V. Reboud[4], V. Calvo[3], J. Widiez[4], J.M. Hartmann[4], A. Chelnokov[4], J. Faist[5] and H. Sigg[1+]

[1] Laboratory for Micro- and Nanotechnology, Paul Scherrer Institut, 5232 Villigen, Switzerland

[2] Univ. Grenoble Alpes, CEA, IRIG, MEM, 38000 Grenoble, France

[3] Univ. Grenoble Alpes, CEA, Grenoble INP, IRIG, PHELIQS, 38000 Grenoble, France

[4] Univ. Grenoble Alpes, CEA, LETI, 38054 Grenoble, France

[5] Institute for Quantum Electronics, ETH Zürich, 8093 Zürich, Switzerland

+email: hans.sigg@psi.ch; francesco.armand-pilon@psi.ch

## Abstract

Efficient and cost-effective Si-compatible lasers are a long standing-wish of the optoelectronic industry. In principle, there are two options. For many applications, lasers based on III-V compounds provide compelling solutions, even if the integration is complex and therefore costly. However, where low costs and also high integration density are crucial, group-IV-based lasers - made of Ge and GeSn, for example - could be an alternative, provided their performance can be improved. Such progresses will come with better materials but also with the development of a profounder understanding of their optical properties. In this work, we demonstrate, using Ge microbridges with strain up to 6.6%, a powerful method for determining the population inversion gain and the material and optical losses of group IV lasers. This is made by deriving the values for the injection carrier densities and the cavity losses from the measurement of the change of the refractive index and the mode linewidth, respectively. We observe a laser threshold consistent with optical gain and material loss values obtained from a tight binding calculation. Lasing in Ge – at steady-state - is found to be limited to low temperatures in a narrow regime of tensile strain at the crossover to the direct band gap bandstructure. We explain this observation by parasitic intervalence band absorption that increases rapidly with higher injection densities and temperature. N-doping seems to reduce the material loss at low excitation but does not extend the lasing regime. We also discuss the impact of the optically inactive carriers in the L-valley on the linewidth of group IV lasers.

INTRODUCTION

Today, germanium-based materials, strained [1] or alloyed with Sn [2 3], form the most advanced platform for group-IV lasing. However, despite encouraging results [4 5 6 7 8 9 10 11 12 13 14], all-group-IV lasing performances are behind those of group III-V based lasers integrated on silicon using direct growth on prepatterned [15 16] or planar layers of quantum-dots [17 18], quantum-wells [19] or quantum – cascade structures [20]. There is definitely the possibility to close this gap, by improving the materials and strengthening the knowledge of the fundamental mechanisms relevant for lasing in group-IV. For the latter, we miss a coherent description of the dependence of lasing performances on the band offset and doping. It is also essential to understand the role of parasitic losses, such as the intervalence band (IVB) absorption [21] and its dependence on the carrier injection and temperature. Moreover, even essential details of the band structure, such as the band offset between the band edges at $\Gamma$ and L and its dependence on the strain, are not fully settled yet, neither for Ge [22] nor for GeSn [23] alloys. In order to improve the performance of group IV laser, such would have to be known precisely.

Here, we address these points for the strained Ge system where we can benefit from thoroughly developed tools to calculate bandstructure [24] and properties relevant for transport [25] and lasing [26 27]. Moreover, Ge is attractive not only for laser benchmarking, but also because of its recent use in quantum technology as a spin qubit [28, 29, 30]. The possibility to fabricate from the same material quantum gates as well as a laser source provides further motivation for this study.

The bottleneck to overcome to achieve lasing action with pristine Ge is the low population of carriers at the centre of the Brillouin zone ($\Gamma$) where optical transitions are allowed. Being an indirect semiconductor, Ge differs from standard direct band gap semiconductors such as GaAs, where basically only such $\Gamma$ states are occupied. There are 2 approaches to enhance the carrier population at $\Gamma$ in Ge. (i) The initial proposal suggested to use strong n-doping with phosphorous [31]. However, the related experiments showed contradictory results. Whereas the photo-[32] and electro-luminescence [33, 34] from optical cavities made of such highly doped Ge had been attributed to lasing, the absorption studied under optical pumping on similar material in Ref [35, 36] did not reveal the net gain required for lasing. Alternatively, by applying tensile strain, (ii) we can control the population at the $\Gamma$-point by gradually closing the offset between the bandedge at the $\Gamma-$ and L-points until we have reached and even passed the cross over to direct bandgap, where – in a situation similar as the one of GaAs - the lowest energy states in $\Gamma$ are below the ones of L, and population inversion between the relevant states and thus lasing becomes possible.

In order to systematically investigate this transition regime from an indirect to a direct semiconductor, we make use of the recently developed strain amplification approach [37 38]. It enables to study the evolution of lasing – using the same doped or undoped base material – as a function of just the strain. Lasing under tensile strain has also been investigated for the GeSn system [9, 12], however its dependence on the strain has not been systematically studied.

The organization of the paper follows our experimental approach: we first describe the strain amplification concept [37, 38] used to produce a series of high quality optical cavities with strain up to 4.2% at room temperature, reaching up to 6.6 % at low temperature. For all samples, the low temperature photoluminescence spectra of the cavity modes are investigated in dependence on the excitation power, i.e. the carrier density. The latter was determined from the plasma shift of the dielectric function. We infer the strain from matching the experimental band gap to the calculated bandstructure obtained using the tight binding (TB) approach [24]. With the same model we compute the matrix elements needed for the interband gain and the intra valence band material loss calculation. We show that by choosing appropriate values for the carrier temperature and line broadening, as well as the cavity losses, the TB-model accurately predicts lasing thresholds for samples with approximately 6% of strain for excitation around $2 \cdot 10^{18}$ cm$^{-3}$. The model also explains the observed lasing roll-over at higher excitations and when the cryostat temperature exceeds 30 K. Via the analysis of the cavity modes' linewidth - similarly to the work of Petykiewicz et al. [39] - we deduce the cavity quality factor and we gather an overall understanding of the gain and loss mechanism in our tensile strained Ge cavities. We then pin down the crossover to direct bandgap for uniaxially strained Ge at 6.1%. In the final experimental sections, we study the mode linewidth of moderately n-type doped material with high strain and discuss a peculiar deviation of the usual Schawlow-Townes laser linewidth theory [40] for the case of only modestly direct bandgap group IV lasers.

## I. SAMPLES

In this experiment, we study the optical properties at low temperatures of highly strained Ge microbridges which are suspended and embedded into an optical cavity. The high strain is achieved by patterning along <100> and selectively under-etching a GeOI layer with a built-in biaxial tensile strain $\varepsilon_0$ of 0.16%. The layers were undoped or doped in-situ by phosphorous. The strain amplification method has been explained in detail in [37, 1, 41]. We employed samples with notations L5, L6, L7 on chip A and L9, L10, L11 on chip B obtained from two consecutive processing runs. The labels indicate structures with progressively increasing total pad length L (in μm), as shown in Fig.1(a). The length L determines the uniaxial tensile strain, which ranges from 3.7 up to 4.2 % at room temperature (RT) as measured via Raman scattering. The strained microbridge of length $l_b$ = 8 μm is integrated in a strain-maintaining optical cavity with two corner cube reflectors, at a distance $l_c$ = 44 μm. The cavity is shown in Fig.1(b), together with the optical mode pattern as calculated with Comsol Multi Physics.

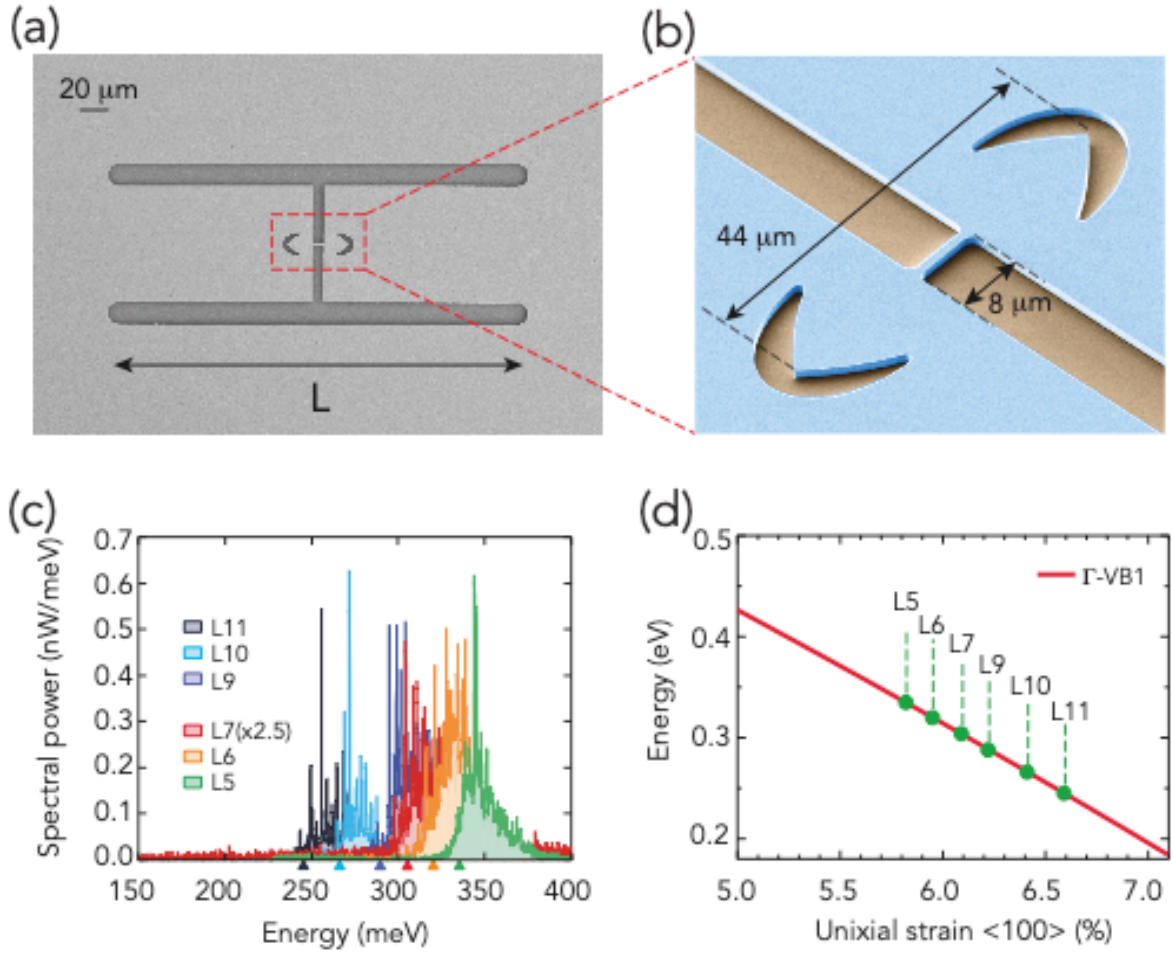

FIG.1. *(a)* Top-view Scanning Electron Microscope (SEM) image of a strained germanium microbridge with a total length of L. *(b)* Tilted SEM image in false colour of a corner cube cavity. Superimposed, the Finite Element Method (FEM) simulation of the out of plane electric field of the fundamental TE cavity mode. *(c)* Samples L5 to L11 photoluminescence at 15 K upon

The microbridge structures are mounted in a cryostat with a base temperature of 15 K and are excited with a diode laser running in continuous wave with an energy of $h\nu_{ext}$ = 590 meV ($\lambda$ = 2.15 µm) and an excitation diameter (i.e. a full width at half maximum) of about 8.7 µm. The photon energy of the excitation is well below the bandgap of unstrained Ge to guarantee low loss in the pad region which is essentially unstrained. Throughout this paper, the reported excitation power is defined as the total power at the sample position. Owing to the excellent thermal conductivity of Ge at low temperature (1.7 kW/m·K at 15 K), the temperature increase of the samples is less than 10 K at the excitation of 10 mW, even when considering a possible reduction of the thermal conductivity by one order of magnitude due to the structure's finite size. The optical spectra are recorded via a Fourier transform infrared spectrometer (Bruker Vertex 70) running in the fast scanning mode with an instrumental resolution of 0.5 cm$^{-1}$, corresponding to 62 µeV.

The low temperature photoluminescence (PL) spectra of samples L5 to L11 are reported in Fig.1(c). Each of the samples show a broad spontaneous emission strongly modulated by Fabry-Perot cavity modes. The higher the strain of the sample, the lower is the PL energy. Compared to RT, - c.f. Appendix A - the PL onset is red shifted, in line with a tensile strain increase when cooling the samples [41]. We converted the measured PL signal to spectral power of light collected by the microscope using the calibration procedure described in [1]. This procedure could not be applied for L7, because of a misaligned detector. Its spectral power was adjusted by an arbitrary factor of 2.5 for better visibility in Fig. 1(c). Fig.1(d) reports the Γ - VB1 bandgap energy dependence on strain at low temperature as obtained from the tight-binding (TB) simulations [24, 1]. As per Fig.1(d), we deduce the strain of the samples from the position of the band gap. This approach was verified previously by Guilloy et al. [22] via electro-modulation spectroscopy on similar microbridges with strain up to 3.3 % calibrated by x-ray microdiffraction [42]. Here, we relate the bandgap to the photon energy at the half maximum spectral intensity of the photoluminescence background, c.f. x-axis of Fig.1(c). This simple method is particularly well suited in case of strong cavity modes. Corresponding strain values are summarized in Table I.

|  | L5 | L6 | L7/L7* | L9 | L10 | L11 |
| --- | --- | --- | --- | --- | --- | --- |
| Band gap (meV) | 335 | 320 | 304 | 288 | 266 | 245 |
| Strain (%) | 5.82 | 5.95 | 6.09 | 6.22 | 6.41 | 6.59 |

*TABLE I. Band gap energy extracted from the photoluminescence at 15 K and the deduced strain value (c.f. Fig.1(b) and (c)).*

From the correlation of the PL spectra with Raman measurements taken at room temperature, c.f. Appendix A, we obtain that the thus defined bandgap values fit nicely the strain dependence of the TB model up to 4.2%.

## II. CAVITY SPECTRA AND LASING THRESHOLD

Fig.2(a) to (e) show a series of power dependent cavity-PL spectra taken at 15 K for samples L5, L6, and L7*, L9, L10 and L11. The sample denoted as L7* belongs to chip B, has the geometry of L9 but the PL spectra characteristics and onset of the L7 cavities of chip A, c.f. Appendix B. We may thus safely assume that the strain of that particular sample L9, henceforth referred to as L7*, is as for the L7. To enable measurements under identical conditions on a series of differently strained samples, we use in the following the results of L7* together with L9, L10 and L11.

Samples are excited up to 15 mW, corresponding to a carrier density of about $3.7 \cdot 10^{18}$ cm$^{-3}$. This conversion is based on the observed frequency shift of the cavity modes induced by the electron-hole plasma dispersion effect, as shown in Appendix C. The factor of $0.25 \cdot 10^{18}$ cm$^{-3}$/mW is found to be valid for all samples. This value corresponds roughly to a case with an absorbed fraction and a recombination time of 40% and 5 ns, respectively. The latter is in agreement with the surface recombination time found in a previous pump probe study of the 2 μm thick GeOI source material used here [43] while the 40% is a fair estimate for strained Ge when considering the cavity effect created by multiple reflection within the bridge and via the Si substrate.

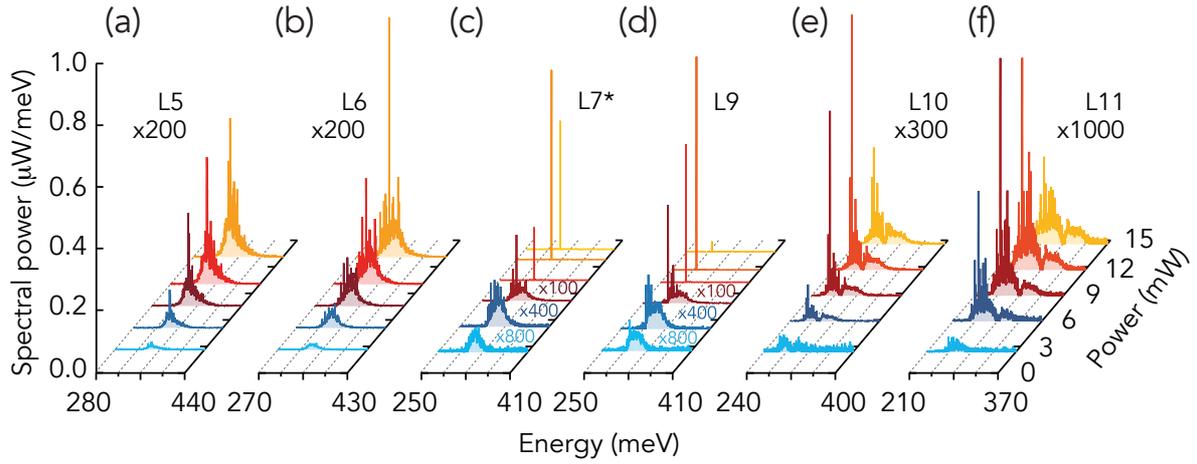

**FIG.2.** *PL emission power collected at the exit of the microscope versus continuous wave excitation power taken at 15 K of (a) L5, (b) L6, (c) L7\*, (d) L9, (e) L10 and (f) L11 microbridge cavities. Above threshold, the L7\* and L9 spectral emissions of (c) and (d) are dominated by a single mode at 311 and 295 meV, respectively.*

The spectra shown in Fig.2 can be grouped in 3 strain ranges. For the lowest strains, Fig. 2(a) and (b), the spectra show a broad photoluminescence background modulated by Fabry-Perot cavity modes. For intermediate strains, the spectral emission of L7\* and L9 in Fig. 2(c) and (d) show a similar combination of background and multi-mode emission, but only up to about 6 mW excitation. Above about 9 mW, there is a single mode peak in the PL spectra of L7\* and L9: its spectral peak intensity sharply increases by about two to three orders of magnitude above the spontaneous emission background. Remarkably, for the highest strains, Fig. 2(e) and (f), the spectra of L10 and L11 are again multi-mode up to the highest excitation power.

The emission obtained for the samples with intermediate strains not only exhibit single mode but also reveal a steep increase of the total emitted power, as shown in Fig. 3(a), clearly demonstrating lasing action in L7\* and L9. Conversely, L5 and L6 as well as L10 and L11, the samples with the lowest and the highest strains, do not lase, as neither single mode operation nor a noticeable increase of emission efficiency are observed.

From the intersect between the linearly extrapolated PL emission at high and low excitation power shown in Fig. 3(a), threshold powers of 9.6 and 8.6 mW for respectively L7\* - as well as L7 shown in Appendix B - and L9 are obtained. These values, using the above mentioned conversion factor, correspond to a carrier density $N_{th}$ = 2.4 (2.1)·$10^{18}$ cm$^{-3}$ for L7*(L9). For both samples, the linear increase in the emitted intensity is followed by a striking roll over. For L9, the roll over point occurs at an excitation power lower than for L7*.

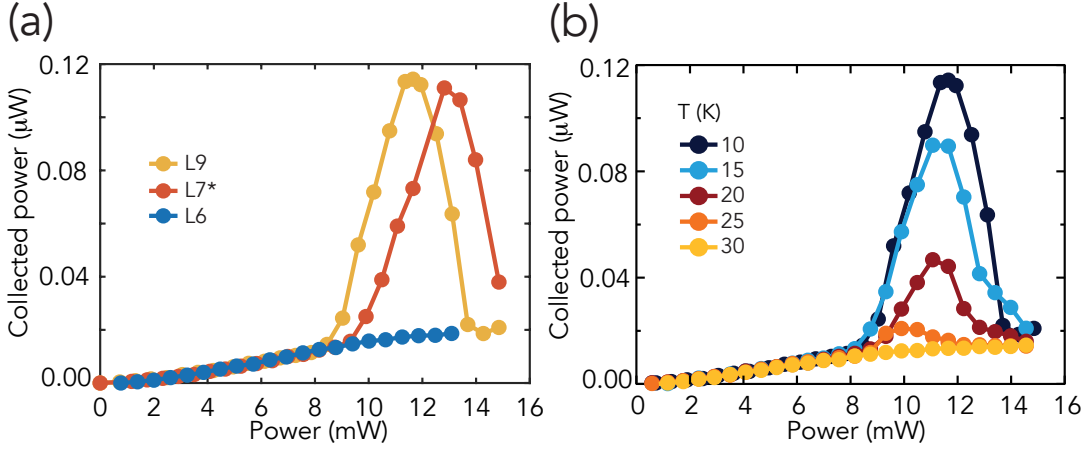

*FIG.3. (a) PL power at the exit of the microscope as a function of the excitation power for samples L6, L7\* and L9 at 15 K. (b) Collected power of sample L9 for different temperatures, from 15 up to 35 K.*

In Fig.3(b), we show the light-in light-out curves of lasing sample L9 as a function of temperature in linear scale. The temperature is increased in steps of 5 K from 15 K up to 35 K. By raising the temperature, the lasing efficiency rapidly degrades until at 35 K the lasing is suppressed. This behaviour, together with the roll over at higher excitation as well as the threshold values of L7* and L9 near 2·$10^{18}$ cm$^{-3}$ are conform to the predictions detailed in the following.

### III. BANDSTRUCTURE, GAIN AND LOSS

Fig. 4(a) represents the bandstructure of germanium under a uniaxially loaded tensile strain of 6% (shown in orange) when ΔE, the offset between the Γ and the L band edges, is practically zero. The blue lines represent the bandstructure for relaxed Ge when the strain is zero and ΔE is about 140 meV. Details of the used tight binding model are given in Ref [24]. The strongest optical transitions are indicated by the green and red arrows representing, respectively, the interband transitions providing gain and the intervalence band (IVB) transitions responsible for the main material loss, $\alpha_m$. Fig. 4(b) shows gain spectra for the barely direct bandgap case with ΔE = - 3 meV and carrier density up to 5·$10^{18}$ cm$^{-3}$. We give further details about the optical gain model in Appendix D.

Carrier temperature and state broadening for this calculation are set to 30 K and 10 meV, respectively. The latter comes from a previous IVB absorption study of doped Ge [44] at low temperature. As no corresponding study was made for the interband excitation, its state broadening is assumed to be the same for simplicity. The carrier temperature is expected to be higher than the base temperature set for the cryostat and possibly also above that of the phonon bath, because carriers can be heated by the injection and might be out of equilibrium, albeit only weakly. However, as a change of the base temperature by only 20 K – as demonstrated in Fig. 3(b) – is able to switch of the lasing, we know that the heating of the carries by injection is low and will not exceed 30 K at the 15 K base temperature, as we will conclude later.

The open circles shown in Fig 4(b) mark the peak gain for a selection of excitation densities. While the peak gain is shifting to higher energies with increasing excitation, a plateau with zero absorption develops at the high energy side of the spectra. This behaviour is typical for marginally direct systems; it can be attributed to the filling of the lowest valence band states near $\Gamma$ by a large number of holes, while most of the excited electrons are in the L-conduction band. Fig. 4(c) shows the IVB loss spectra ($\alpha_m$) for several excitation densities assuming the same carrier temperature and state broadening. Curves are tagged by open squares at the photon energy corresponding to the peak gain at the same carrier density. The IVB absorption strongly increases for decreasing energies and its high energy tail is characterized by a rapid increase for carrier densities and temperatures above $2 \cdot 10^{18}$ cm$^{-3}$ and 30 K, respectively. Detail of our IVB absorption model is presented in Appendix E together with an analysis of its dependence on temperature and state broadening. The rapid increase of the IVB absorption is due to holes filling the states further apart from the $\Gamma$ point in reciprocal space, c.f. Fig. 4(a). The strongest contribution comes from the transitions between the VB1 and the VB2, as indicated by the red arrow in Fig. 4(a). The transitions from VB1 to the split-off (VB3) occur at energies larger than 650 meV and thus are not relevant in the present case.

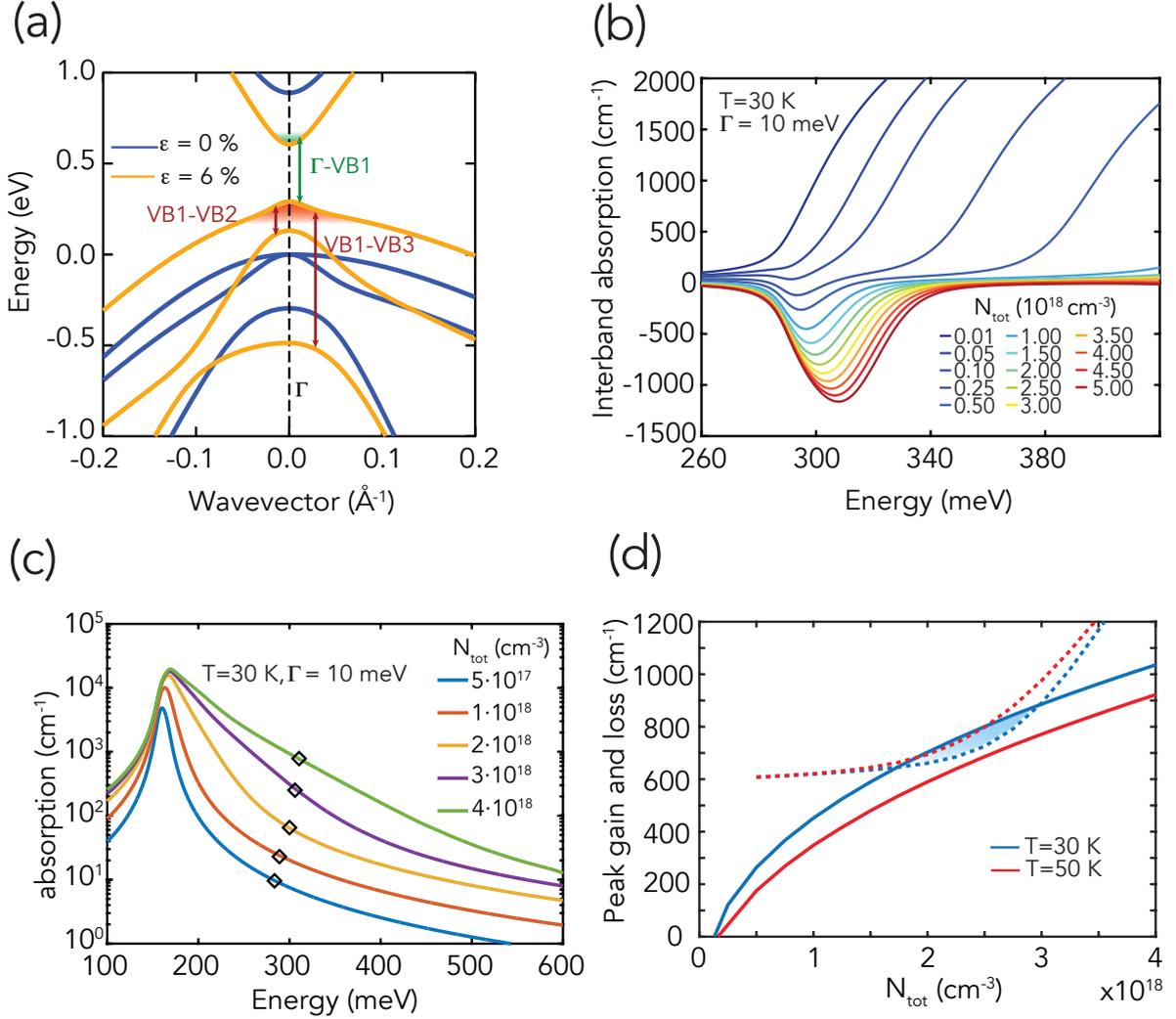

*FIG. 4:* Modelling of Ge's band structure, gain and loss. **(a)** Band structure of germanium near the $\Gamma$ point, calculated with the tight-binding model for strain along the [001] direction of 6 % (orange) when $\Delta E=0$, and for 0 % (blue) when $\Delta E=140$ meV. The green and red arrows indicate the interband and intervalence band transitions. **(b)** Interband absorption spectra for carrier excitation up to $5 \cdot 10^{18}$ cm$^{-3}$ for $\Delta E = -3$ meV, with a bandgap about 290 meV. A transparency region develops above the gain region. **(c)** Intervalence band absorption spectra for 6 % of strain and excitation densities of 0.5, 1, 2, 3, and $4 \cdot 10^{18}$ cm$^{-3}$. In **(b)** and **(c)** the electron temperature and state broadening are assumed to be 30 K and 10 meV, respectively. **(d)** Peak optical gain (solid lines) as a function of the injection density for an offset $\Delta E = -3$ meV, extracted from **(b)** (c.f. black circles). The total loss (dashed lines) in function of the carrier density is obtained by adding the intervalence band absorption calculated at peak gain energy (c.f. open square in **(c)**) on top of the optical loss of 600 cm$^{-1}$. The latter value is obtained from the experiment and is normalized to the mode filling of 0.17, see main text. The blue and red colour represent the carrier temperatures of 30 and 50 K, respectively. For 30 K, the lasing threshold is found to be near $2 \cdot 10^{18}$ cm$^{-3}$ while at higher densities (> $2.5 \cdot 10^{18}$ cm$^{-3}$) a laser roll-over occurs because of the steeply rising IVB absorption. When the carrier temperature is set at 50 K, the gain (red, solid line) does not overcome the loss (red, dashed line).

Fig 4(d) shows the dependence on the excitation density of the peak gain (blue curve) and the corresponding loss (blue dashed curve) at 30 K. Lasing is expected when the modal gain overcomes the total loss which includes material and optical or cavity loss, $\alpha_c$. The latter is experimentally found to be 103 cm$^{-1}$ as detailed in the next section. The material loss can be extracted from Fig. 4(c). It has to be evaluated at the emission energy where the gain is maximal, as indicated by the open squares We show in Fig. 4(d) the case of a marginally direct bandstructure, with an offset $\Delta E$ = -3 meV. For this case, the gain overcomes the total loss at an injection density of about $2 \cdot 10^{18}$ cm$^{-3}$. For zero offset (not shown), the gain does not overcome the IVB loss anymore, demonstrating that the offset $\Delta E$ is the main parameter driving the gain at the low temperatures probed here. Optical broadening and effective carrier temperature also have some impact on gain and loss, but only marginally. We observe that, for any combination of optical broadening and effective temperature variation in the range of ± 5 meV and ± 15 K, respectively, the offset at which the gain may overcome the loss shifts by less than ± 2 meV. However, we reach the best overall agreement with the experiments with the above values of broadening and temperature.

Due to the fast-growing IVB absorption with density and temperature, lasing is achieved neither at higher excitation nor elevated temperature. As shown in Fig. 4(d), the low temperature loss curve (blue dashed line) overcomes the gain at around $3 \cdot 10^{18}$ cm$^{-3}$. At 50 K, the gain (red solid line) does not prevail over the loss (red dashed line). Both predictions are in good agreement with the experiments presented in Fig. 3(a) and (b).

In fact, by considering in our gain calculation the offset ΔE as free parameter and shifting the gain spectrum to the experimental PL onset, the observed lasing threshold is reproduced for ΔE = - 1.5 meV (L7*) and – 2.5 meV (L9).

We will show below, that we can understand with our model – keeping the values of the parameter the same - the lack of lasing in L10 and L11, but only when considering that the strain dependence of the directness is smaller than anticipated from the TB model [24].

## IV. LINEWIDTH ANALYSES AND CAVITY LOSS

In Fig.5(a), we show the linewidth evolution with respect to the excitation power of the L9 cavity modes labelled in Fig.5(b). The reported linewidths Δ(hν) are apodized. We identify three different regimes: (i) The linewidth of modes 1 to 4 narrows because of the gain. In line with the assumption that lasing occurs at the gain peak, the lasing mode, labeled as 1, narrows the most. (ii) The linewidths of the high energy modes 5 and 6 are largely independent of the input power. Their values scatter around 0.45 meV. We relate this effect to transparency, as outlined in the next paragraph. (iii) Near and above lasing threshold, the linewidth levels off. The non lasing modes from 2 to 4 saturate at 0.20 meV. This behavior can be well understood from the Fermi level pinning when lasing occurs. (iv) Unexpectedly, also the lasing mode 1 does not narrow but saturates at about 0.15 meV, well above the spectral resolution of our experiment. In principle, floating nanocavities tend to vibrate, which could periodically shift the cavity's resonant position. Such a behavior leads to a peaked noise spectrum [45]. However, the noise spectrum observed for L7* when lasing, is found to be flat. We thus attribute the finite laser linewidth to refractive index fluctuation induced by the carriers in the L-valley and multimode lasing on nearly degenerate modes, c.f. the hereafter following section VII. *Laser linewidth*.

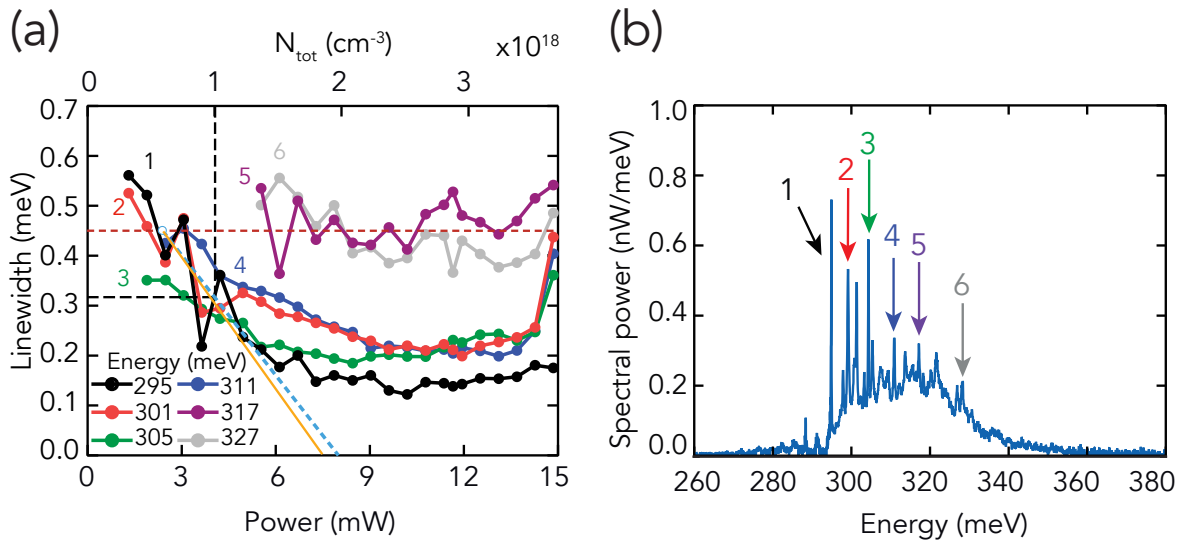

**FIG.5. (a)** *L9 linewidth at 15 K as a function of the excitation power of the modes labelled in **(b)**. The horizontal dashed line at 0.45 meV indicates transparency. The differential net gain is extracted from the linear regression between the transparency point and the power corresponding to $N_r=10^{18}$ cm$^{-3}$. It is shown as a dashed light blue line. The orange line starting as well from transparency is obtained from the model of Fig.4(b) and (c). **(b)** L9 spectral emission at 15 K for an excitation power P=6.12 mW.*

Via the observation of (ii), we gain access to cavity loss. In line with our Fig. 4(b) gain calculation, a wide transparency band (i.e. zero absorption) occurs for transitions with energies above the gain region. We make use of this characteristic behavior and translate the 0.45 meV linewidth of modes 5 and 6 into a cavity loss $\alpha_c$. We obtain from $\Delta \nu = 1/(2\pi\tau)$, where $\tau$ is the photon decaying time, equal to $\tau = n_g/(c\alpha_c)$, where $n_g$ is the group refractive index, $\alpha_c \approx 103$ cm$^{-1}$. This procedure is well supported by the observation that all samples, save the highest strained sample L11, have high energy modes' linewidths saturating at about 0.45 meV. Intrinsic material losses, such as the parasitic free carrier and IVB absorption, are largely negligible at high energies, in line with our model.

From the standard threshold condition, $exp(\Gamma_{xy}g_{th}l_b)exp(-\alpha_c l_c)exp(-\Gamma_{xy}\alpha_m l_b) = 1$ where $\Gamma_{xy}$ = 0.94 is the transversal confinement factor, $g_{th}$ the material gain at threshold, $l_b$ = 8 μm the active material length and $l_c$ = 44 μm the length of the cavity, we can deduce that the active material delivers at threshold a gain of: $g_{th} = \Gamma_{xy}^{-1} \cdot \alpha_c\, l_c/l_b + \alpha_m$.

The gain required to start lasing is thus quite significant due to the low filling factor $\Gamma_{xy} \cdot l_b/l_c = 0.17$ of the active material. Even without any material loss, the necessary gain for lasing is thus 600 cm$^{-1}$.

The differential gain is obtained via the slope of the linewidth narrowing starting from transparency up to the excitation power corresponding to a reference density of N$_r$ = 10$^{18}$ cm$^{-3}$. Up to this density, the IVB absorption increase is linear. We consider the mode showing the strongest reduction, which is obviously the lasing mode for L7* and L9. For the latter, this is shown as a dashed light-blue line in Fig.5(a). We then convert the slope to the differential net gain using the linewidth definition in unit of energy $\Delta(h\nu) = hc\alpha/(2\pi n_g)$. For L9, we find dg/dN = 4.33·10$^{-16}$ cm$^2$, which is close indeed to the calculated value of g(N$_r$)/(N$_r$ − N$_0$) = 4.70·10$^{-16}$ cm$^2$, as obtained for the offset value of ΔE = -2.5 meV derived by correlating the calculated to the measured N$_{th}$. N$_0$ corresponds to the carrier density at transparency. Moreover, for L7*, the offset value determined from the linewidth/gain analyses is in close agreement with ΔE obtained from threshold density analyses. These conformities give confidence for the slope based offset determination of the non-lasing samples, L5, L6, L10 and L11. Values are given in Table II together with the deduced linewidth reduction per mW of excitation. The corresponding linewidth analyses for L9, L10 and L11 are detailed in Appendix F, together with the linewidth excitation dependence obtained for these samples at T = 30 K. To circumvent possible issues because the gain may increase non-linearly near transparency, in particular for L10 and L11, the excitation density corresponding to a reference linewidth value of L9 is used for an alternative offset determination. Again, the resulting offsets – c.f. Table II - are in close agreement with the ones based on the slope method. For an overview of the three methods to obtain the ΔE from the threshold and/or the experimental linewidth reduction, we refer the reader to Appendix G.

| Method \ Samples | Undoped | | | | | |
|---|---|---|---|---|---|---|
| | L5 | L6 | L7* | L9 | L10 | L11 |
| LD | | | -1.5 meV | -2.5 meV | | |
| DL (measurement) | 3.0 meV (-38 meV/W) | 2.0 meV (-47 meV/W) | -1.0 meV (-70 meV/W) | -2.0 meV (-80 meV/W) | -4.0 meV (-96 meV/W) | -6.0 meV (-116 meV/W) |
| LC | | | | | -4.0 meV | -5.0 meV |
| TB model | 4.8 meV | 1.7 meV | -2.3 meV | -4.9 meV | -9.7 meV | -14.5 meV |

| Method \ Samples | Doped | |
|---|---|---|
| | Low: $1.8 \cdot 10^{18}$ cm$^{-3}$ | High: $6.3 \cdot 10^{18}$ cm$^{-3}$ |
| DL (measurement) | -6.5 meV (-133 meV/W) | < -10 meV < (- 200 meV/W) |

*TABLE II. Band offset ΔE of undoped and phosphorus doped samples obtained from the laser threshold (LD), the differential linewidth (DL) and the linewidth comparison (LC) to L9 as outlined in the main text, in comparison to tight binding (TB) model calculations. The differential linewidths obtained from the experiments shown in Fig. 5, Fig. 7, and Fig. 14 are included in brackets.*

While the experimental offset values agree among each other, the comparison to the model is striking. Firstly, by interpolation of ΔE between L6 and L11, we obtain that the crossover occurs for a uniaxial tensile strain of 6.05 %. This is in excellent match with the model. In contrast, the dependence of the offset on strain differs considerably: we obtain the experimental slope is smaller than predicted by almost a factor of two. Possibly, the slope determination may be (i) affected by the simplifications made in the modelling of gain and IVB absorption, or (ii) impacted by technical issues, such as material quality imperfection or bridge distortion that may introduce a non-uniaxial component of stress. We address some of these points in Appendix I. In principle, of course, it would also be possible (iii), that the theory is inaccurate with respect to the slope. To prove this, the offset would have to be measured not only near the crossing at around 6% of strain, but also at medium strain, about 3%. Photo-reflection measurements [46] could be the most suitable method here. Despite these unresolved discrepancies between experiment and model, we are confident about the determination of the crossover, since it is based on the observation of lasing at low temperatures. The crossover determination by means of the analysis of the temperature dependence of the PL intensity - as presented in Ref [47] for the case of strained Ge- thus turns out to be inaccurate in retrospect.

## V. N-DOPING

As next we study the PL of strained microbridge cavities made of the in-situ doped Ge, with phosphorous (P) concentrations of 1.8 and $6.3 \cdot 10^{18}$ cm$^{-3}$. We refer to these samples as low doping (LD) and high doping (HD) sample, respectively. The strain, as inferred from the onset of the PL, is about 5.8 %, similar to that of the undoped microbridge sample L6. PL results are shown in Fig.6(a) and (b). The doped samples do not lase. However, compared to the spectra of L6 shown in Fig.2, cavity modes already develop at low power and the lines narrow fast with the excitation power, c.f. Fig. 6(c). However, as the linewidth saturation also sets in already at low power, both doped samples seem far from reaching the lasing threshold. Lasing seems even further away than for L6, as found by comparing the LD and HD spectra at highest power in Fig.6(a) and (b) and the corresponding one shown in Fig 2(b).

The linewidth slope method yields an offset for both doped samples which is negative, namely -6.5 and -10 meV, respectively. In fact, for the higher doped sample, the slope shown in Fig. 6(c) is only a lower limit, as the linewidth was already of 0.3 meV at 2 mW excitation. Thus, compared to the 2.0 meV offset for L6, phosphorous doping seems to reduce $\Delta E$ by about 8.5 meV and more than 12 meV for the LD and HD samples, respectively. Interestingly, these numbers are in reasonable agreement with the Fermi level positions in the L conduction band of the LD and HD samples, namely 8.4 meV and 20 meV, respectively.

We thus seem to have obtained by n-doping a system that becomes quasi direct, as we determine offset values which are negative. However, for higher excitation, the gain does not develop further and thus does not overcome the optical loss, particularly in the high doping case The most likely explanation for the reduction of the gain at high excitation is a strong reduction of the lifetime caused by the Auger effect. This explanation is consistent with the observation of the positions of the cavity modes which shift more and more slowly for the doped samples at increasingly higher excitation density. Alternatively, the gain may not be pure interband but relate to donors-to-valence band transitions that change their character from L to Γ when the two conduction bands align. Such band tail states [48] may give a weak but insufficient gain.

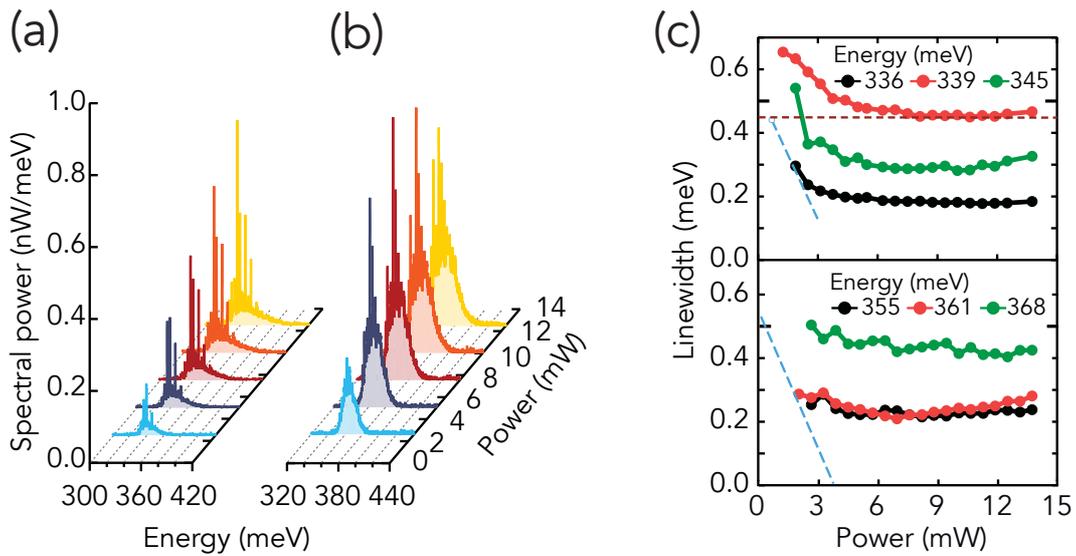

**FIG.6.** Power dependent photoluminescence spectra at 15 K of equally strained Ge:P cavities with **(a)** $N_d=1.8 \cdot 10^{18}$ cm$^{-3}$ and **(b)** $N_d=6.3 \cdot 10^{18}$ cm$^{-3}$. **(c)** up and down panel compare the linewidth dependence on the excitation power of the LD and HD samples, respectively. The dashed lines represent the linewidth slope used for the $\Delta E$ determination.

## VI. LASER LINEWIDTH

For the characterization of the laser linewidth, we focus on sample L7*. The laser resonance of L9 overlaps with an absorption band in air which complicates the linewidth analyses. Examples of the measured laser line spectra of L7* are shown in Appendix H, together with the fit function that consist of Lorentzian broadened lines.

For the discussion, we follow the standard model of Schawlow Townes (S-T) [40], which describes the linewidth of a single mode laser above threshold as: $\Delta\nu = \eta_{ST}(1 + \alpha^2)/P_{out}$, where $P_{out}$ is the radiated power per round trip. The factor $\eta_{ST}$ depends on the amount of gain at threshold, the cavity loss and the degree of population inversion [40]. The linewidth enhancement factor α describes the coupling between intensity and phase noise [49]. The $\Delta\nu$ value is notably governed by the fact that above threshold, gain clamping fixes the amount of carrier density to $N_{th}$, thereby stabilizing the spontaneous emission and thus the random drift of the phase.

Fig.7 top panel shows the measured linewidth – below and above threshold - as a function of the excitation power. On the lower panel, the laser linewidth is show as a function of the inverted collected power. When lasing starts the linewidth is inversely proportional to the output power as predicted by the theory. In contrast to theory, however, the experimental linewidth remains finite, and approaches a residual value $\Delta\nu_0$ of about 21 GHz, i.e., 0.09 meV, whose interpretation we give in the next paragraph. Before, we estimate the $\alpha$ value from the linewidth dependence on the power by converting the collected power to the total output radiated power $P_{out}$, c.f. upper scale of the Fig.7 lower panel. The conversion is calculated from the corner cube far field emission pattern as explained in [1]. We thus obtain that the linear increase is reproduced by the S-T model for the case of α=0 and the parasitic loss as shown in Fig. 4(c). A low value for α means that the change in carrier density does not change the absorption but the peak gain. This is not exactly what we find from the gain calculation shown in Fig.4(b) which result in an α value of about 1.7. This value is indeed smaller than what is found for truly direct band gap systems, such as III-V compounds, where α values are between 2 and

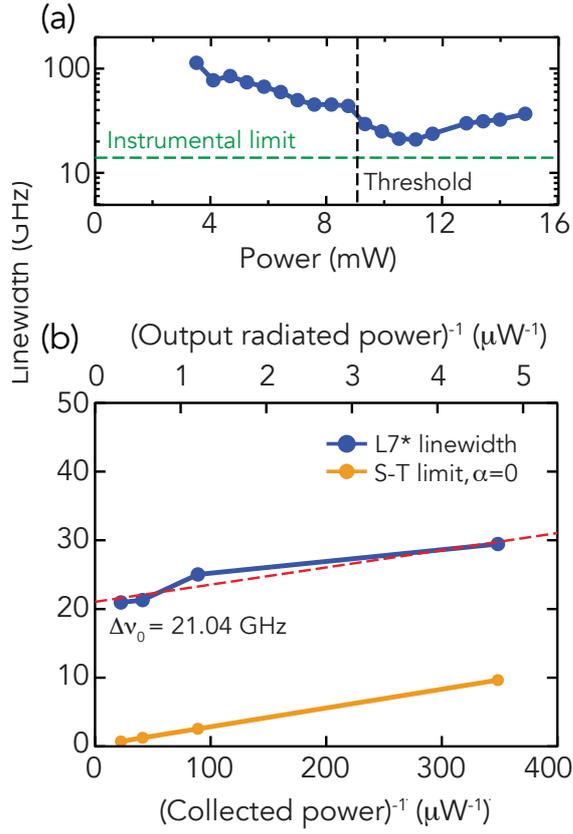

FIG.7. Experimental linewidth of the L7* lasing mode (blue). **(a)** linewidth as a function of excitation power. The instrumental resolution at 15 GHz is indicated in green. **(b)** L7* linewidth as a function of the inverse of the collected power when the sample is operated above threshold but before the roll over. The red dashed line is a linear fit to the experiment, yielding a residual linewidth $\Delta \nu_0$ of about 21 GHz. In yellow: linewidth of the standard model of Schawlow Townes for $\alpha = 0$ as obtained from the estimated radiated output power given on the upper scale. It is deduced from the collected power as described in the text.

6. Considering the uncertainty in the determination of the power conversion factor as well as the simplification in the gain modelling, the agreement is reasonable.

While different experiments validated the Schawlow-Townes theory [50, 51, 52] it was often observed that the linewidth of semiconductor lasers was limited by a power-independent contribution $\Delta\nu_0$ [53, 54, 55, 56, 57, 58, 59]. The origin of the power independent contribution is still debated. It is often attributed to noise, which adds on top of the quantum one, like 1/f noise [60, 61] or occupation fluctuation noise [62]. However, in those cases the linewidth saturates at values in the range of 1 – 10 MHz, orders of magnitude lower than in our case.

A much stronger broadening mechanism is refractive index fluctuations induced by carrier's number noise. This effect, usually neglected thanks to gain clamping, is relevant in injection-controlled tuneable-wavelength lasers, as studied in [63, 64]. We apply this carrier noise model to the strained germanium laser, where most of the carriers are not optically active because they populate the L valley and are thus not clamped by gain. From [63, 64] adapted to our case, we obtain a $\Delta\nu_0$ value of 3.6 GHz, c.f. Appendix H. The remaining quantitative discrepancy may be attributed to multimode operation, which is not included in the S-T model. Indeed, we observe a line doublet, c.f. Appendix H, separated in frequency by about 26 GHz.

## VII. DISCUSSION

The most remarkable observation of the current investigation concerns the existence of a sweet spot of strain at which lasing occurs. This surprising experimental result can be reproduced by the model as demonstrated in Fig.8(a) which gives the peak gains as functions of total carrier densities for samples L6, L9 and L11. For the calculation of the gain, we used the experimental values for ΔE reported in Table II. Superimposed, we show the total loss due to cavity loss and intervalence band absorption. As before, the IVB absorption is calculated for the energy where the gain peaks. We thus obtain that L5 and L6 do not reach the threshold due to the low amount of delivered gain and the fast rise of IVB absorption for densities larger than 2 to $3·10^{18}$cm$^{-3}$. In contrast, the gain increases fast enough in L9 (and also L7 as well as L7*) to reach the lasing threshold at about $2·10^{18}$ cm$^{-3}$. The onset of the IVB absorption explains the observed break up of lasing at high excitation shown in Fig.4(a). Finally, L11 does not reach the threshold because strong parasitic loss occurs just before the gain is large enough to overcome the optical losses. As shown in Fig. 4(d), the temperature dependence of the lasing is – at least qualitatively – well explained by the model with phenomenological temperature and broadening parameters. Fig. 8(b) shows that the experiment and theory are definitely in line concerning the crossing to a direct band gap configuration, but not in the slope, which is shallower in the experiment. We have argued that our method is more accurate for determining the cross-over than for determining the slope.

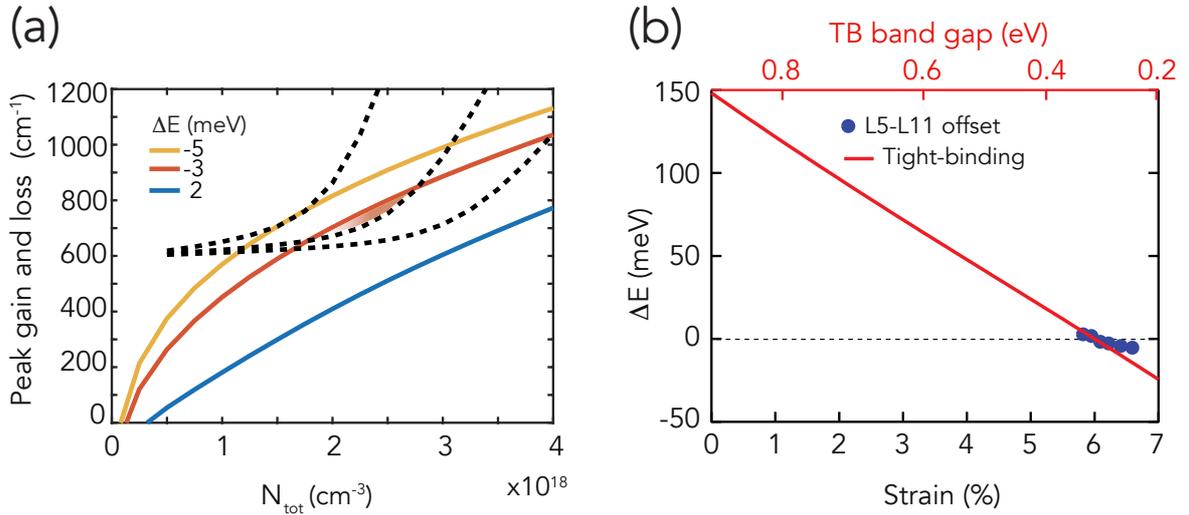

**FIG.8. (a)** *In color, peak optical gain as a function of the total carrier density up to $4·10^{18}$ cm$^{-3}$ for the approximated energy difference ΔE of L6, L9 and L11. For L9 and L11 we choose the more accurate values obtained with the threshold and linewidth reference methods, respectively (c.f. Appendix G). The overall loss due to cavity loss and parasitic inter-valence band absorption is shown as a black dashed line. It was extracted from Fig. 4(c). **(b)** Energy difference ΔE as a function of uniaxial strain at low temperature. In red the prediction from the tight binding theory, in blue dots the experimental points extracted from Table II. The upper axis converts via the tight-binding model the strain to the Γ -VB1 transition energy.*

The indirect to direct crossover determined here also confirms the results from Ref [1], where samples, similar to those investigated here, strained up to 5.9%, thus with indirect bandgap were studied. Samples of Ref [1] showed lasing at low temperature, but only upon pulsed excitation, with a pulse length of 100 ps. This was attributed to (i) indirect bandgap, i.e., $\Delta E > 0$, and (ii) the build-up of a transient population inversion, thanks to the blocking of the phonon-mediated $\Gamma$ to L intervalley scattering. Such process is particularly effective for electrons with excess energy lower 28 meV, which corresponds to the energy of the zone-boundary longitudinal acoustic phonon. As a matter of fact, the offset $\Delta E$ of the least strained samples showing lasing and the most heavily loaded samples showing no lasing of Ref [1] are indeed less than 20 meV and about 28 meV, according to the TB calculation for strain of 5.4 % and almost 5%, respectively.

## VIII. CONCLUSION

We developed a versitale laser analysis method for determining the gain and the material and optical losses of group-IV lasers. It is made by deriving the values for the injection carrier densities and the cavity losses from the measurement of the refractive index change and the mode linewidth, respectively. We exemplified the method using Ge microbridges strained up to 6.6%.

We determined the strain required to lift germanium across the transition from an indirect to a direct band structure by comparing the calculated gain and loss with experimental observations of laser threshold and/or differential gain. We found a sweet spot for lasing threshold for uniaxial tensile strain of approximately 6.1%, which also marks the crossover to a direct bandgap semiconductor. At lower strain, the optical gain is too low to overcome cavity losses, while samples at higher strain suffer from IVB absorption, which critically depends on the transition energy, excitation strength and temperature. The obtained crossover is well predicted by the tight-binding model [24] and confirms the transiently blocking of the valley scattering model of Ref [1]. The experimental slope by which the offset changes is, however, smaller than predicted. The adoption of an optimized cavity, like a distributed Bragg reflector as in [65], might extend the strain and temperature ranges where lasing is achieved, enabling future studies to address whether the $\Delta E$ disagreement is of fundamental or experimental nature.

Our method revealed that doping brings a slight gain enhancement, but only at low excitation, which we attribute to (i) additional gain from doping induced band tail states and (ii) a reduced lifetime due to Auger scattering, respectively.

We discussed the fundamental physics of the laser linewidth in the case when carriers fill dominantly the L levels which are not optically active and thus are not clamped by gain. We expect that such L-carrier fluctuations may also play a role in other group IV lasers, depending on the actual offset and the temperature of the laser.

Altogether, we have shown a practical way to achieve lasing in Ge at low temperature and in steady state. We related the lasing to the crossover to a direct bandgap. Our analyses method can perfectly be applied to laser cavities realized with other material systems such as GeSn alloys, hexagonally ordered SiGe [66], and defective Ge [67] and thus could greatly accelerate the future development of group IV lasers for the integration into Si microelectronics.

ACKNOWLEDGMENTS

This work was supported by the CEA-Grenoble program Phare Photonics, and by the Swiss National Science Foundation SNF. The authors acknowledge the assistance of the staff of CEA Leti clean rooms and of CEA Advanced technological platform PTA and acknowledge technical support for the experiments performed at PSI from Stefan Stutz.

**Appendix**

## APPENDIX A: ROOM TEMPERATURE PL INVESTIGATION

Fig.9(a) compares the experimental values of the Γ-VB1 transition energy at room temperature with models' predictions for various strain levels. The red and green colours give the value of the tight-binding (TB) [1, 24] and the deformation potential [68] models, respectively. The solid and dashed lined refer to the Γ-VB1 and L-VB1 transitions, respectively. Predictions are compared with experimental results for samples with strains determined by Raman scattering. Black circles show the values inferred from the work of Guilloy et al.[22], showing a good agreement up to 3.3 % with the TB prediction of the Γ-VB1 transition. The validity of the TB model is further confirmed up to 4.2 % by the values highlighted in blue, extracted from the photoluminescence measurements of samples L9, L10 and L1, as shown in Fig.9(b).

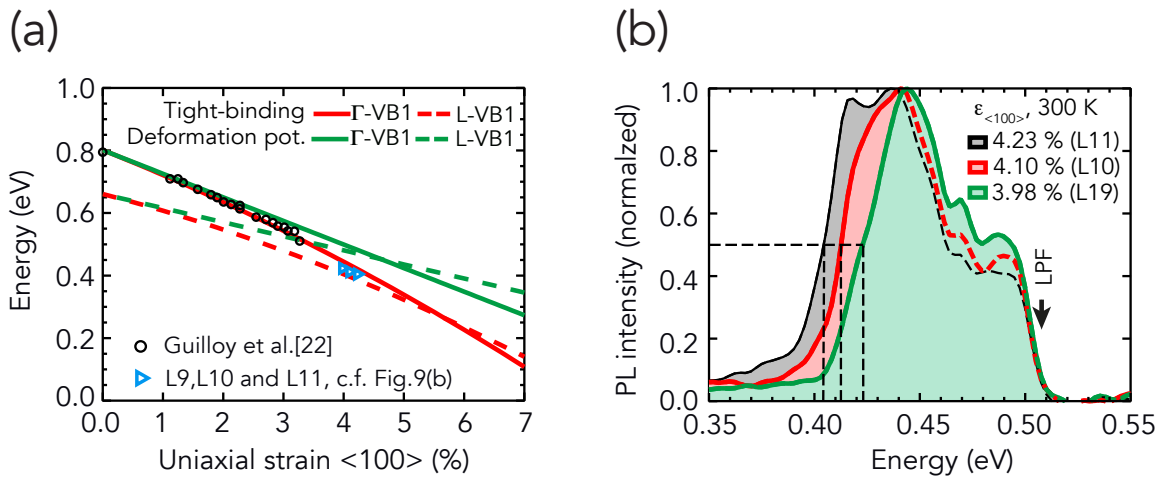

*FIG.9. (a) Interband transition energies in Ge, for uniaxial strain along one of the equivalent crystallographic directions <100> at 300 K. Solid lines refer to the Γ-VB1 transition, while the dashed lines refer to the indirect transition L-VB1. Energies from the tight-binding and deformations potential models are in red and green, respectively. Experimental values are inferred from [22]. (b) low resolution photoluminescence measurement at 300 K upon continuous wave excitation running at 590 meV of samples L9, L10 and L11. The high energy PL is blocked by a lowpass filter (LPF).*

Samples L9, L10 and L11 are strained at 3.98, 4.10 and 4.23 %, respectively. As for the measurements taken at low temperature, the band gap position is estimated as the energy corresponding to the half maximum spectral intensity of the photoluminescence background, c.f. Fig.9(b). The spectral features between 450 and 490 meV are attributed to absorption in air. The emission is suppressed at high energy by a long pass filter cutting below (above) 2440 nm wavelength (508 meV). The low energy side of the photoluminescence of Fig.9(b) is impacted by the tail of the thermal background.

# APPENDIX B: Comparison between sample L7 and L7*

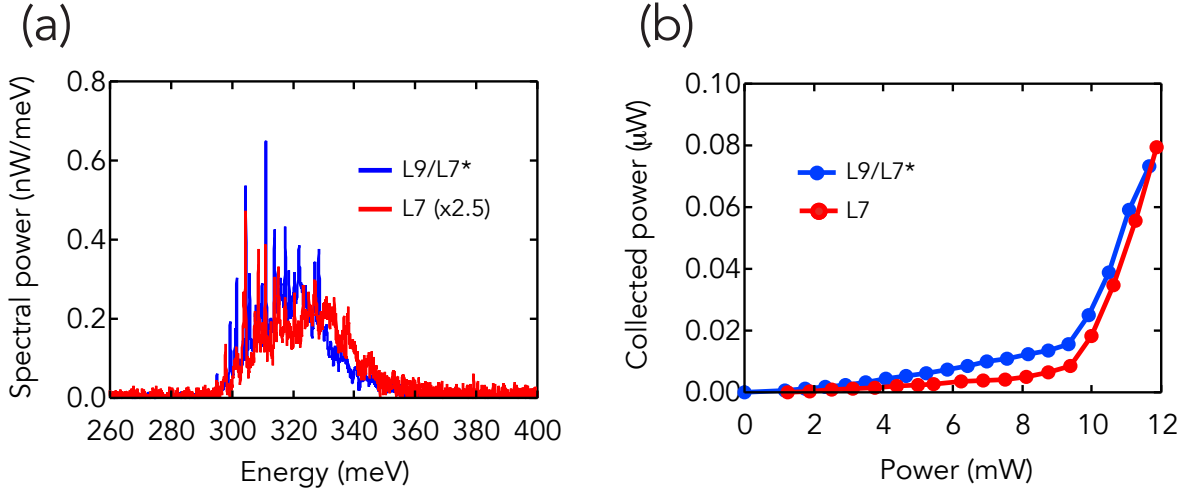

**FIG.10. (a)** *PL spectra of samples L7 (in red) from chip A and a selected L9 (in blue) from chip B, referred to as L7\* in the main text taken at 15 K and an excitation power of about 6.3 and 6.4 mW, respectively. The spectrum of L7 is enhanced by factor 2.5 for comparison.* **(b)** *PL power as a function of the excitation power at 15 K.*

Fig.10 (a) shows PL spectra of an L7 sample from chip A and a selected L9 sample from chip B with excitation intensity is below the laser threshold. Except for the intensity - which we attribute to a detector misalignment for the case of sample L7 - the spectra look very similar with an onset at the same energy of about 300 meV. The latter indicates that both the samples are loaded with approximately the same strain. Moreover, considering that all the investigated samples have identical optical cavities and thus differ only by their amount of strain, the equal laser threshold, c.f. Fig. 10 (b), further indicates the match between this sample pair. Therefore, to enable measurement sequences on the same chip B not only for samples L9, L10 and L11, we decided to include also this specific L9 cavity in our study, and denote it as L7*.

# APPENDIX C: CAVITY MODE SHIFT

The refractive index change Δn caused by the injected carrier N is described in the Drude based model [69] as:

$$\Delta n = -\frac{q^2 \mu_0 \hbar^2 c^2}{2n_g (h\nu_{mode})^2 m_0 \mu_{plasma}} N \quad (B1)$$

where $h\nu_{mode}$ is the photon energy of the considered cavity mode, $\mu_0$ the vacuum permeability, $n_g$ the unperturbed material group refractive index of 4.5, and $\mu_{plasma}$ is the plasma mass of 0.06. We calculate the energy shift of the cavity mode from the relation $\Delta h\nu/h\nu = -\Delta n/n$ and by considering transversal and longitudinal fundamental mode confinement factors of $\Gamma_{xy}$ =0.94 and $\Gamma_z$ =$l_b/l_c$=8/44, respectively.

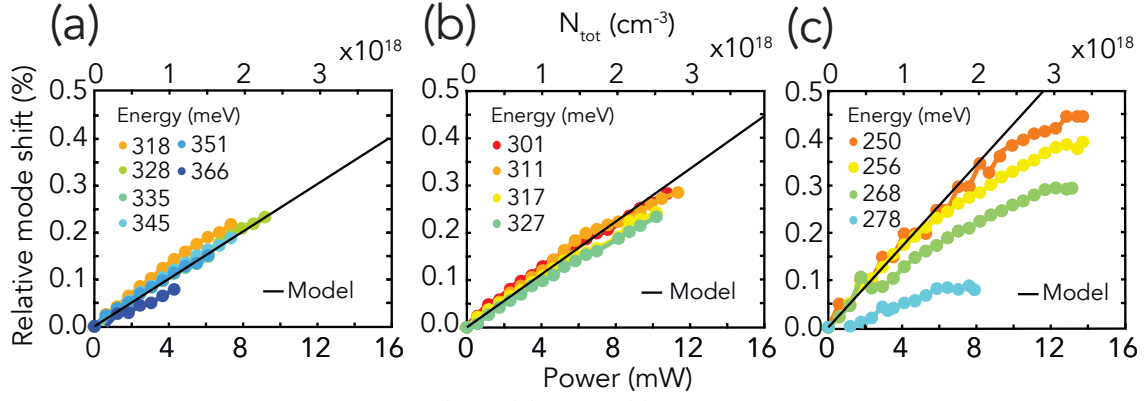

**FIG.11.** *Relative mode shift $\Delta h\nu/h\nu$ of samples **(a)** L6, **(b)** L7\* and **(c)** L11 at 15 K in function of the excitation power. The carrier density, given on the upper scale, is calculated from the mode shift according to the Drude model, shown by the black line.*

In Fig.11 we report the cavity modes shifts at 15 K of L6, L7* and L11 as functions of the excitation power. On the upper scale we give the carrier concentration as calculated from the Drude model of Eq. (B1). It appears that the thus obtained conversion factor of $0.25 \cdot 10^{18}$ cm$^{-3}$/mW is roughly the same in all the investigated samples, independent of the temperature (between 15 and 30 K). For the in-situ phosphorous-doped samples (not shown here), the match is only at low carrier densities.

## APPENDIX D: Optical gain modelling

The absorption coefficient is calculated as the ratio between the net photon number absorbed per second per volume unit and the photon number incident per second per area unit. By means of the Fermi's golden rule, we calculate the net rate of photon absorption per unit of volume. We then obtain the following expression integrated in the reciprocal space for the absorption of the photon energy $\hbar\omega$ [26]:

$$\alpha(\hbar\omega) = \frac{\pi q^2}{n_r c \varepsilon_0 m_0^2 \omega} \int \frac{2\, d^3k}{(2\pi)^3}\ |\hat{e}\cdot p_{if}|^2\ \delta(E_f - E_i - \hbar\omega)[f_v(E_i) - f_c(E_f)] \quad (C1)$$

where $E_i$ and $E_f$ are initial and final energies of the inter band transition. The carrier statistics is described by the Fermi distribution function:

$$f(E, T, \mu) = \frac{1}{exp\left(\frac{E-\mu}{k_b T}\right) + 1} \quad (C2)$$

where µ is the quasi-Fermi level, and by the three-dimensional density of states:

$$\rho^{3D}(E, E_0, m^*) = \frac{1}{2\pi} \frac{(2m^*)^{(3/2)}}{\hbar^3} \sqrt{E - E_0} \quad (C3)$$

The quasi-Fermi levels are obtained self-consistently by inverting the charge neutrality equation: $N_{el}$ = $N_h$ = $N_{tot}$ and assuming thermal equilibrium at temperature T. The electron and hole densities are obtained by integration over both conduction bands (L and Γ) and the three valence bands, respectively:

$$N_{tot}^{el} = \int_{\min(E_\Gamma, E_L)}^{\infty} \sum_{i=\Gamma, L} \rho_i^{3D}(E, E_i, m_i^*) f_c(E, T, \mu_c) dE \qquad (C4)$$

and

$$N_{tot}^{h} = \int_{E_{HH}}^{-\infty} \sum_{i=VB1, VB2, VB3} -\rho_i^{3D}(-E, -E_i, m_i^*) [1 - f_v(E, T, \mu_v)] dE \qquad (C5)$$

Table III gives the effective masses and the band edges calculated by means of the tight-binding model for 6 % of strain, corresponding to ΔE = 0 meV. $m^l$ is the component longitudinal to the strain, i.e. along [001], while $m^t$ is the component along [100] and [010]. For the L band, $m^l$ and $m^t$ are the component along the [111] and [1-10] directions, respectively.

| Band | $E_i$ (eV) @20 K | $m_i^l$ | $m_i^t$ |
|---|---|---|---|
| Γ | 0.314 | 0.03329 | 0.01813 |
| L | 0.314 | 1.55567 | 0.09067 |
| VB1 | 0 | 0.17987 | 0.02763 |
| VB2 | -0.160 | 0.12504 | 0.04192 |
| VB3 | -0.777 | 0.03715 | 0.23103 |

*TABLE III. Longitudinal and transverse masses at 6 % of uniaxial tensile strain along [001] for the Γ and L bands and for the three valence bands VB1, VB2 and VB3. The values are calculated using the tight-binding model and are expressed in units of $m_0$*

Table IV gives the dipole matrix elements calculated from the tight-binding model for 6 % of strain. The selection rules make no distinction between light TE and TM polarized propagating along the strain direction.

| Transition | $\mid \hat{e} \cdot p_{if} \mid^2$ |
|---|---|
| Γ–VB1 | $(m_0/2) * 11.32$ eV |
| Γ–VB2 | $(m_0/2) * 10.65$ eV |
| Γ–VB3 | $(m_0/2) * 0.61$ eV |

*TABLE IV. Dipole matrix elements for transition between the Γ valley and the valence bands for radiation polarized perpendicularly to the strain, calculated with the tight-binding model for 6 % of uniaxial tensile strain.*

The following approximations were used for the gain calculation: (i) for all strain, we used the dipole matrix element and the masses as given above. (ii) the band dispersion was treated as quadratic because the probed range of **k** around the Γ point is small. (iii) Electron-electron and electron-phonon processes were not considered and (iv) thermal equilibrium was assumed.

Fig.12 shows the contour map of the peak gain as a function of the total carrier density and energy offset ΔE, calculated using the formulation described above. The effective carrier temperature is set to 30 K. The gain spectrum is convoluted with a Lorentzian function with a full width half maximum of Γ=10 meV, as explained in the main text. When germanium becomes direct, the number of carriers available for optical recombination in Γ, and thus the optical gain, increase strongly, making the gain an excellent probe of the bandstructure's directness.

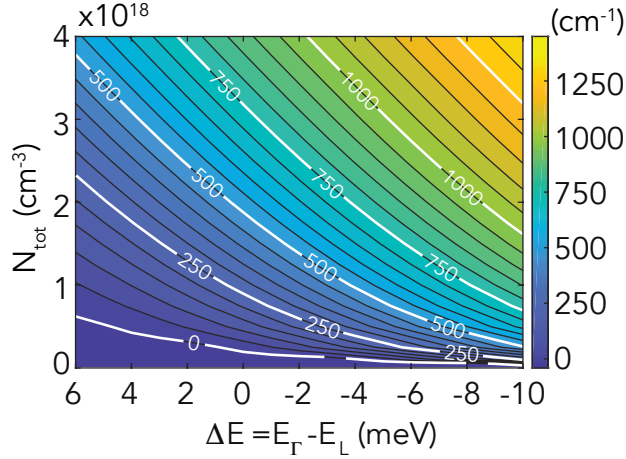

**FIG.12.** Contour plot of the optical peak gain for a carrier temperature of 30 K and an optical broadening of 10 meV, as a function of the energy offset ΔE and total carrier density, $N_{tot}$.

## APPENDIX E: INTERVALENCE BAND ABSORPTION

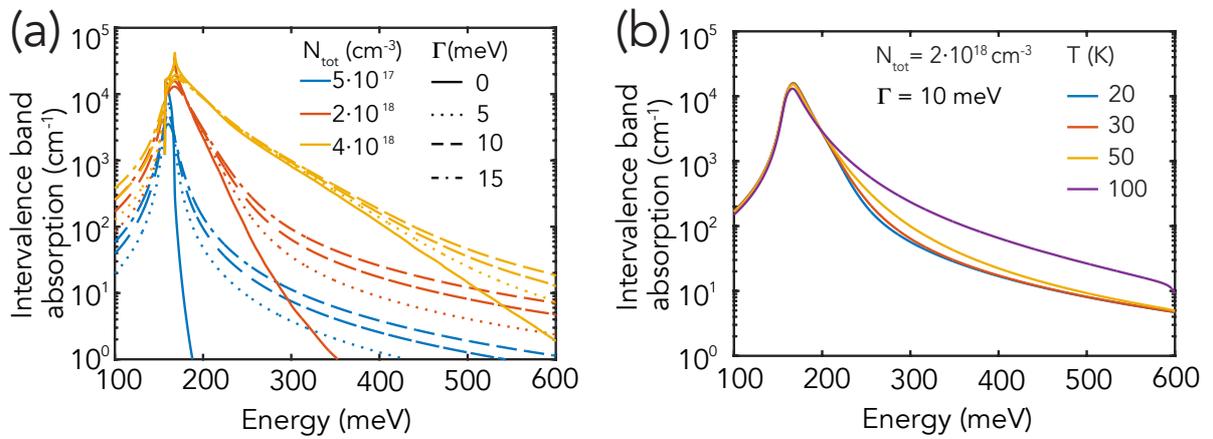

**FIG.13.** Intervalence band absorption spectra calculated with the tight-binding model for 6 % of strain, corresponding to zero energy difference ΔE, **(a)** at 30 K for different carrier densities (color) and state broadening (line): solid, dotted, dashed and dash-dotted correspond to 0, 5, 10 and 15 meV broadening, respectively and **(b)** at $N_{tot}$=2·10$^{18}$ cm$^{-3}$ and state broadening of 10 meV for different temperatures. Calculations consider only TE and TM polarized light, propagating along the direction of Fig.13(a) and (b) report the IVB absorption spectra for different carrier density, state broadening and temperature between 20 and 100 K. Spectra are calculated with the tight-binding model for 6 % of strain [24]. The tight-binding valence band structure and inter-valence band dipole matrix elements are fed into Eq. (C1), and the integration over wave vector is performed on a tetrahedral mesh around $\Gamma$. The transition between the upper and the lowest valence bands are neglected because they do not contribute at the here relevant strain and transition energies. Included, in contrast to the interband case, however, is the non-parabolic bandstructure and the k-dependence of the matrix elements.

We note that the IVB absorption strongly increases for decreasing energies and peaks at about 160 meV. Moreover, its high energy tail increases for carrier density above $2 \cdot 10^{18}$ cm$^{-3}$ and temperature T > 30 K. The latter effect is due to holes filling the states in reciprocal space further apart from the $\Gamma$ point, c.f. Fig. 4(a), main text.

The characteristics of the IVB absorption enable us to interpret qualitatively several of the reported effects (main text), such as the L7* and L9 intensity roll-overs that shifts to lower powers when the temperature increases. Because the lasing transition energy is lower for L9 than for L7*, the roll-over - when the parasitic loss starts dominating - is reached faster for the former. The higher the strain, the lower the bandgap energy and thus the stronger the absorption will be. The gain increase for L10 and L11 is thus not enough to reach lasing threshold under the present conditions.

We also like to note that we neglected the free carrier absorption of electrons (FCA). This process annihilates a photon by exciting an electron from an occupied state below the quasi-Fermi level to an empty state above [70]. Since such a process requires an impurity or a phonon to comply with momentum conservation, the absence of phonons at low temperature makes FCA negligible compared to IVB absorption [71]. It can thus be ignored.

## APPENDIX F: L9, L10 AND L11 LINEWIDTH AT 15 AND 30 K

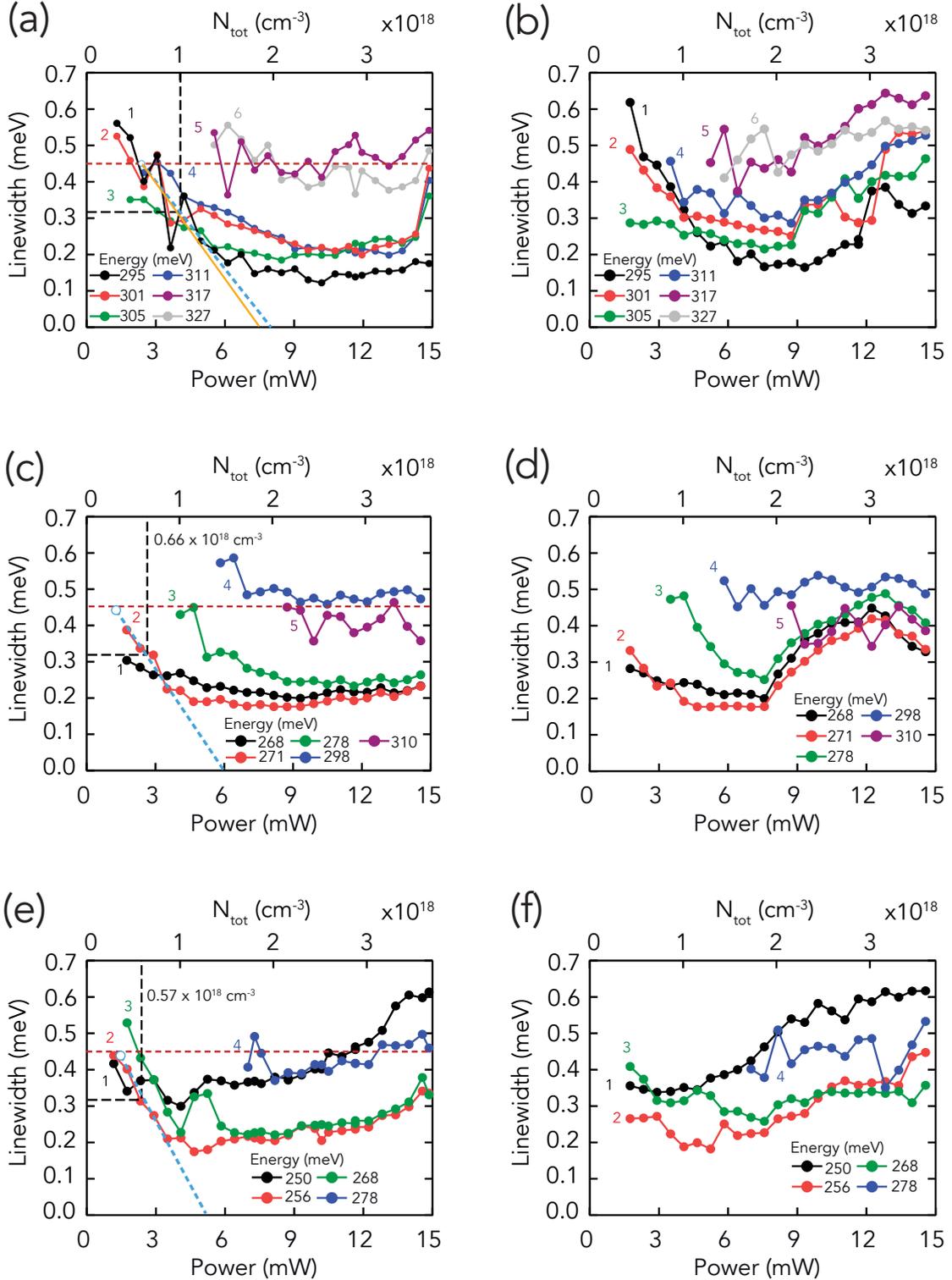

***FIG.14.*** *Linewidth as a function of power and temperature. L9 linewidth at **(a)** 15 K and **(b)** 30 K. L10 linewidth at **(c)** 15 K and **(d)** 30 K. L11 linewidth at **(e)** 15 K and **(f)** 30 K. The black dashed line in **(c)** and **(e)** indicate the carrier density at which the linewidth reaches the value of 0.317 meV, as found for L9 **(a)** at the reference carrier density of $N_r = 1*10^{18}$ cm$^{-3}$ (c.f. Appendix F). The dashed blue line in **(a), (c)** and **(e)** shows the linear regression used to extract the differential gain (see main text)*

Fig.14 shows the apodized cavity mode linewidth of L9, L10 and L11 at 15 K (a, c and e) and 30 K (b, d and f). At the lowest temperature of 15 K, the cavity modes at low energy first narrow because of gain and later, at higher excitation, smoothly broaden. We interpret the transition from narrowing to broadening as the moment the differential loss overcomes the differential gain. The power at which the linewidth starts broadening decreases when reducing the cavity mode energy. For L9, L10 and L11 samples, the onset of broadening occurs at about 12, 9 and 4.5 mW, respectively.

At 30 K the linewidth broadening becomes stronger. Its onset, for L9 and L10 samples, starts at excitation powers of 9 and 7.5 mW, respectively. For the highest strained sample L11, the impact of a temperature increase is less obvious than for L9 and L10. For L11, despite some fluctuations, the onset can be located between 3 and 6 mW, similarly as at 15 K. Comparing the linewidth slope at low power, we observe that gain is largely independent on temperature.

## APPENDIX G: LINEWIDTH to GAIN CONVERSION

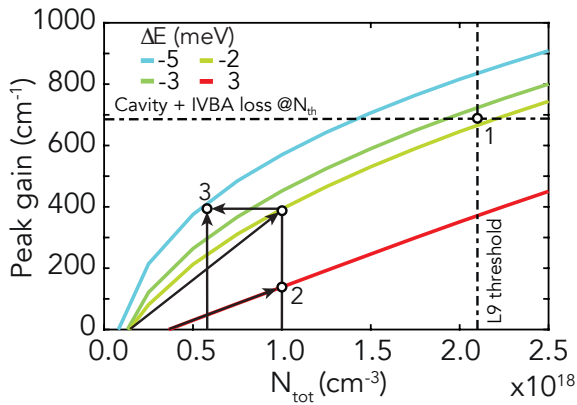

FIG.15. Peak gain as a function of the total carrier density for various energy offsets $\Delta E$. The purpose of this figure is to illustrate the three methods to determine $\Delta E$, as described in the main text.

Fig.15 shows the peak gain calculated as a function of the total carrier density for different offset values $\Delta E$. The three approaches applied in the main text to probe the strained germanium bandstructure are indicated as (1) lasing threshold, (2) linewidth reduction or differential gain, and (3) linewidth or gain reference.

The label (1) in Fig. 15 depicts the situation, for L9, where the gain overcomes the total losses at the lasing threshold density $N_{th}=2.1 \cdot 10^{18}$ cm$^{-3}$. The IVB absorption at the lasing energy of 295 meV and a carrier density of $2.1 \cdot 10^{18}$ cm$^{-3}$ contributes by about 87 cm$^{-1}$ to the total loss of 687 cm$^{-1}$. A corresponding gain is reached for $\Delta E$ between -2 and -3 meV. To probe non-lasing samples, we follow the narrowing of the linewidth from $N_0$ to the reference carrier density $N_r = 1 \cdot 10^{18}$ cm$^{-3}$ (c.f. Fig.5(a) and Fig. 14) and convert this to the differential net gain. $\Delta E$ is obtained by comparing the gain slope to the model, as indicated by label (2), which exemplifies the case of L5. In a first approximation, the IVB absorption at $N_r = 1 \cdot 10^{18}$ cm$^{-3}$ is neglected. For samples L10 and L11, we also compare the carrier density at which their linewidth reaches the value of sample L9 at $N_r = 1 \cdot 10^{18}$ cm$^{-3}$. Parity is obtained for densities of $0.66 \cdot 10^{18}$ cm$^{-3}$ and $0.57 \cdot 10^{18}$ cm$^{-3}$, respectively, c.f. Fig.14(d) and (f). Label (3) in Fig.15 exemplifies the case of L11.

## APPENDIX H: LASER LINEWIDTH OF L7*

Fig.16 shows the L7* non-apodized lasing spectrum for excitation power ranging from 9.90 to 11.07 mW at 15 K. The fitting function suggests the presence of two modes separated by about 0.08 meV. For the linewidth analyzes we have used the linewidth of the strongest mode.

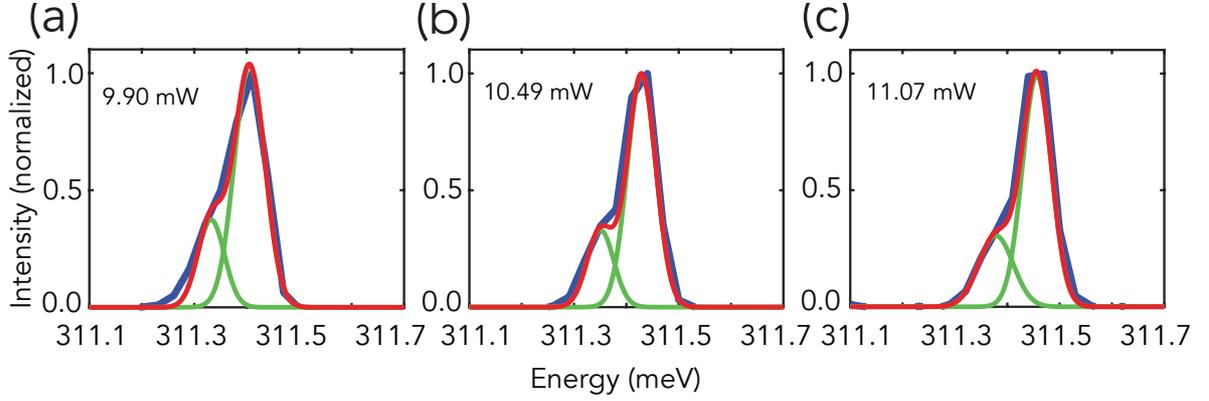

**FIG16.** *(a) In blue the non apodized L7\* lasing spectra at 15 K for excitation power of (a) 9.90 mW, (b) 10.49 mW and (c) 11.07 mW . The fitting function in red is the sum of two Voigt functions (green).*

To evaluate the power-independent contribution $\Delta v_0$ due to noise of the unclamped carrier population in L, we recall the excess linewidth expression developed for the wavelength-tuneable lasers [33], [34]:

$$\Delta v_0 = \frac{q\, I_t}{\pi} \left(\frac{\Delta\omega_{e,mode}\,\tau}{q}\right)^2 \qquad (G1)$$

where $I_t$ expresses the injection tuning current, $\Delta\omega_{e,\,mode}$ is the lasing angular frequency change per unit of charge q, and $\tau$ the electron state lifetime. We apply the above equation to the strained germanium laser case, by expressing $I_t$ as $P_{abs}/(hv_{ext})$, where $P_{abs}$ is - as above - the amount of power absorbed by the microbridge and $hv_{ext}$ the excitation energy of 0.59 eV converted in Joule units. At the threshold of 9.6 mW, $P_{abs} = 0.38\ mW$. We assume that all the carriers populate the L valley. As the change of angular frequency is related to the change of refractive index via the relation $\Delta\omega/\omega = -\Delta n/n$, we can express the change per carrier from the Drude model [69] as already used in Appendix B:

$$\Delta\omega_{e,mode} = -\frac{\omega_{mode}}{n}\frac{\Delta n}{NV} = \omega\frac{q^2\mu_0^2\hbar^2 c^2}{2(n\,\hbar\omega_{mode})^2 m_0 \mu_{plasma}}\frac{1}{V} = 3.39\cdot 10^5\,\frac{rad}{s} \qquad (G2)$$

where N, $\mu_0$, $\mu_{plasma}$ and $n_g$, are as defined in Appendix B. The volume of the strained microbridge V is $8\cdot 10^{-12}$ cm$^{-3}$, while $\hbar\omega$ is the photon energy of 0.311 eV. By inserting equation (G2) in (G1), and by using a carrier lifetime of 5 ns, as discussed in Appendix B, we obtain a linewidth broadening of about $\Delta v_0 = 3.6\ GHz$. From the experiment, we obtain $\Delta v_0 = 21\ GHz$.

The remaining discrepancy is attributed to the fact that our laser does not operate in single mode. For such non-ideal single mode or multimode lasers, the power-independent contribution to the linewidth can reach values above the GHz level due to a non-linear coupling between modes [72].

## APPENDIX I: ADDITIONAL CONSIDERATIONS TO EXPLAIN THE WEAK DEPENDENCE OF THE OFFSET ON STRAIN.

Here we shall consider 3 experimental and technical issues that could explain the shallow evolution of offset energy with strain.

(i) An unintentional misalignment of the bridge structures with respect to <100> would increase $\Delta E$ for a given strain. However, the error margin of our alignment tool, < 1º, is too small to have an impact, making this explanation rather unlikely.

(ii) A similar possibility concerns an inhomogeneous strain distribution as the microbridge structure may bend out of plane. Such bending evokes shear stress which would have an impact on strain. We indeed observe that the expected relationship between the length of the pads and the bridge and their transversal dimensions do not apply for samples with strains larger than in L6. This is why we determined the strain from the measured bandgap energy and did not use its correlation to the geometry, as previously demonstrated [1]. However, shear stress is expected to be minor in the bridge also in case of bending.

(iii) A degradation of the samples' optical properties at the highest strain, for example, due to the generation of a parasitic recombination channel, could result in a gain not evolving as expected. We indeed find that the PL intensity is reduced for samples L10 and L11 compared to the others. However, since – at low excitation strength - the dependence of the charge density on the excitation power is not impacted by strain as obtained from the shift of cavity modes, an upcoming non-radiative recombination path can be excluded.